\documentclass[a4paper,11pt]{article}
\pdfoutput=1 

\usepackage{jinstpub} 

\title{Statistics of the Charge Spectrum of Photo-Multipliers\\ and Methods for
 Absolute Calibration}

\author{Milind V. Diwan}
\affiliation{%
Physics Department, Brookhaven National Laboratory, \\ Upton, NY 11973
}%

\emailAdd{diwan@bnl.gov}

\abstract{
We derive the full expression for the shape of the charge spectrum that results from the illumination of a photo-multiplier tube.  The derivation 
is for low intensity illumination with constant gain, a common condition for most nuclear and particle physics applications.  
 Under these conditions, it is shown that an analytic expression can be formulated that describes  
 details of the spectrum including the pedestal and dark noise with excellent fidelity to allow statistical fits to data.  
The derivation and full formula using either Gaussian or Poisson models for gain,  and its limiting forms under various simplifying assumptions are presented with strategies on their use. The analytic description can be used to formulate data acquisition strategies to perform precise absolute calibration of photo-multipliers, the digitizers, and the data acquisition system.  }

\arxivnumber{1909.05373} 

\usepackage{graphicx}
\usepackage{dcolumn}
\usepackage{bm}

\begin{document}



\maketitle

\flushbottom  


\section{\label{sec:intro}Introduction}
The basic operation of photo-multiplier tubes is explained in many excellent references ~\cite{knoll}. The technology and applications are also explained in user guides from the manufacturers \cite{photonis}.  If illuminated by a small number of photons of the appropriate wavelength, the photo-multiplier cathode will emit  a number electrons depending on the quantum efficiency of the photo-cathode.  Each of these photo-electrons will go  through a multiplication or amplification process through a set of dynodes with total gain that can range from $\sim 10^6 \to 10^7$.  The gain is achieved due to the kinetic energy imparted to the electrons from the high voltage between the cathode and the first dynode, and between each pair of subsequent dynodes.  The usual gain at each dynode ranges  $\sim2 \to 10$ depending on the technology and the applied high voltage.
The gain is assumed to be independent for each photo-electron.  Gain linearity and independence are very general assumptions and are the normal operating conditions in most applications including the some of the largest installations of photo-multiplier tubes for underground neutrino detectors \cite{boger:1999bb,fukuda:2002uc}. 
The total charge is obtain by the multiplication of gains from each stage. For a given average number of incident photons the distribution of the produced charge depends on the statistics of produced photo-electrons as well as the gain, and the addition of noise from various sources.   This has been previously considered in the literature  \cite{bellamy} and used for fitting and extraction of basic detector constants from data.  
We will use some of the conventions from \cite{bellamy}. However,  
in this report, we will not consider the technological aspects of the construction of the photo-multiplier tube, but analyze the statistical properties of the distributions of the total charge produced.  We show that under very general conditions the properties can be described with few simplifying assumptions about the performance of the system.  
In particular, we provide a new general formalism with broad applicability; 
it can be used for a variety of detectors as well as  statistical models for the gain including Poisson fluctuations in the 
number of secondary electrons from the first dynode.  A formula for Poisson gain model is included in the paper. 
Our formula allows  analysis without the assumption of low charge for background processes as assumed in \cite{bellamy}.  
Lastly, we will  analyze the limiting forms of this formula under various conditions commonly in practice. We also point out that 
the  formalism naturally allows a clear understanding of the probability density at or around zero charge which can occur due to the 
sum of fluctuations in charges from backgrounds and signal.  

Although ultimately unnecessary, it may be useful to follow the calculation with a set of assumptions that 
aid visualization of a typical measurement. The photo-multiplier  could be assumed to be subject to a pulsed light source with a low average intensity of photons.  The gain of the photo-multiplier is assumed to be constant and independent for each photo-electron, and the anode where the total charge is collected is considered to be DC coupled to an integrator and/or a waveform digitizer. The total charge and any noise from the tube and electronics is collected in a fixed time interval provided by the expected arrival time of the photon pulse. We should remark that in case of AC coupling, the same considerations remain valid with some modifications to allow for partial collection of the charge, but we will not consider this possibility in detail.

\section{\label{sec:deriv}Derivation } 

The collected charge has three components: the charge from the signal and its multiplication (or amplification), the random noise from the electronics chain, and the noise from  PMT itself in terms of random pulses from the cathode and dynode chain called dark noise. We will build a model for the PMT starting with the background current it produces in the absence of any signal.  

\subsubsection{\label{sec:back}Background Charge} 
\begin{figure}[t]
\includegraphics[width=0.9\textwidth]{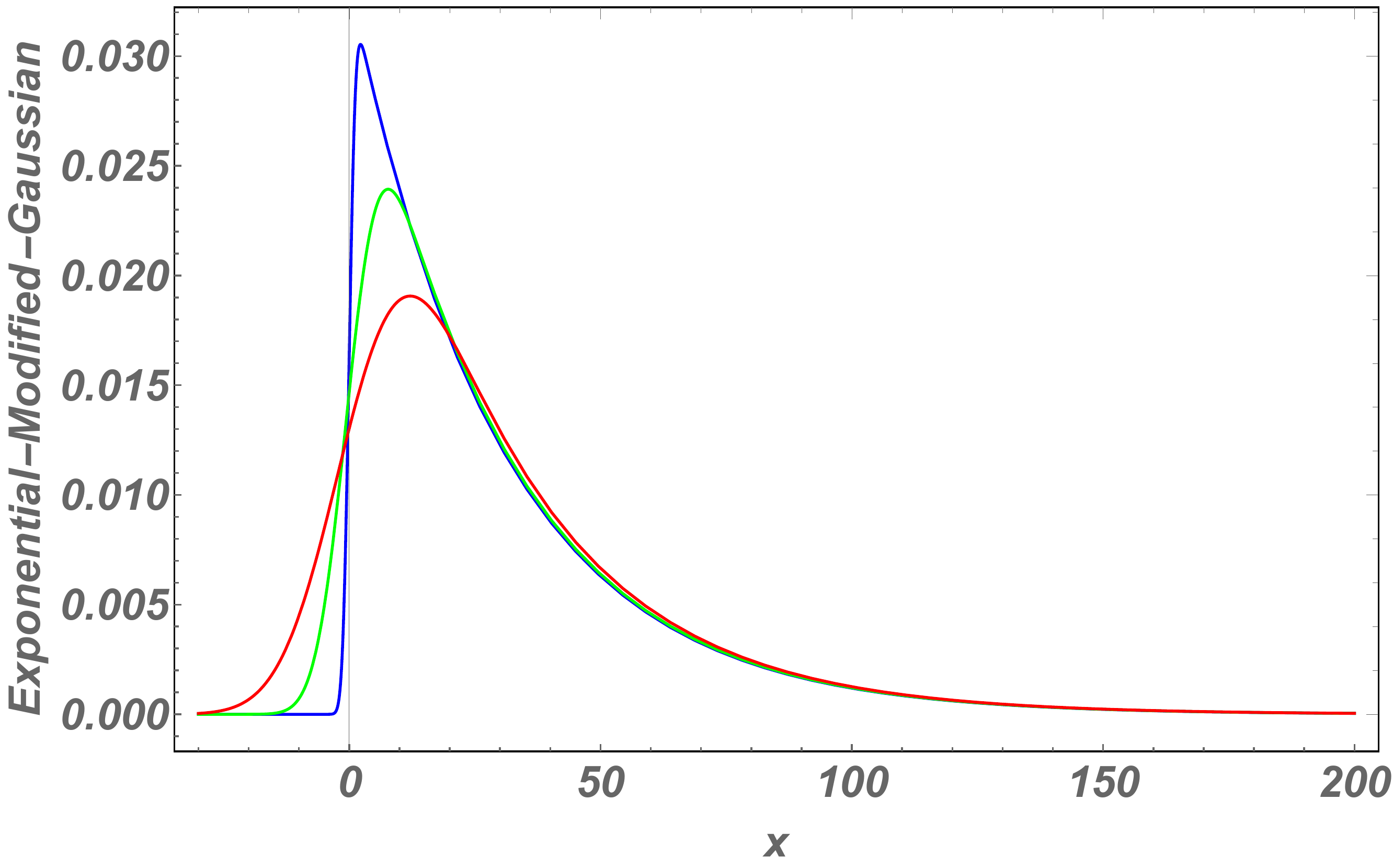}
\caption{\label{fg:emg} 
Examples of the exponentially modified Gaussian (EMG)
probability density function, $E_N(x; \mu, \sigma, c_0)$.  The values of parameters are 
set to be $\mu=0$, $c_0=1/30$, and $\sigma = 1, 5, 10 $ for the blue, green and red curves, respectively.  
}
\end{figure}

The total background is a random number, $B=Q+D $. Here the random number $Q$ is the charge
 due to a fluctuating baseline current without contribution from any electron emission from 
the PMT. We are assuming that the current is integrated over some fixed time interval around the expected arrival of the signal.  
The random number $D$ is the charge due discrete processes  such as thermo-emission from the PMT structures.  
It is obvious that $Q$ will be normally distributed with 
some mean, $q_0$, and standard deviation, $\sigma_0$. These parameters depend on the length of the integration time, but also  on the readout electronics such as the input impedances of the pre-amplifiers and digitizers,  the  background pickup noise, and  the bandwidth of the system.  The background component corresponding to dark pulses, $D$, should  look like single electron emission from the photo-cathode, but that is almost never the case in most practical devices;  $D$ is best described by  an exponential probability density function with an exponential parameter, $c_0$.   The probability distribution functions (PDF) for $Q$ and $D$ are given by: 
\begin{eqnarray}
P_Q(x)=\frac{1}{\sqrt{2\pi}\sigma_0}e^{\frac{-(x-q_0)^2}{2\sigma^2_0}},
\\
P_D(x)=(1-w)\delta(x)+ w\theta(x)c_0 e^{-c_0 x}
\label{eq:backpdf}.
\end{eqnarray}
Here $w$ is the probability of a dark pulse,  $\theta(x)$ is the step function, and $\delta(x)$ is the delta function.  The total background is a convolution of these two probability densities.  The convolution is best calculated by using the characteristic functions for the above PDFs which are given as: 
\begin{eqnarray}
\phi_Q(s)=e^{isq_0}e^{-\frac{1}{2} \sigma^2_0 s^2},
\\
\phi_D(s)=(1-w) + w \frac{1}{1-is/c_0}  
\label{eq:backchar}.
\end{eqnarray}
As a quick reminder,  a characteristic function (CF) is the Fourier transform of a probability density function for continuous random variables or 
a probability mass function (PMF) for discrete random variables. A review can be found from  many resources \cite{pdg}. 
The background PDF can readily be calculated by using the total background characteristic function: 
\begin{eqnarray}
\phi_B(s)= \phi_Q(s)\times\phi_D(s)
\label{eq:bchar}.
\end{eqnarray}
It is easy to see that the total background naturally has two components. The first  term or the pedestal  centered at $q_0$ has a normal PDF, and it is superimposed on the second term,  a somewhat smoothed  exponentially falling dark rate spectrum.  The characteristic function for the smoothed dark rate spectrum is given by:
\begin{equation}
    \phi_{EMG}(s)=\frac{e^{isq_0}e^{-\frac{1}{2} \sigma^2_0 s^2}}{1-is/c_0}
    \label{eq:egchar} 
\end{equation}
This is recognized as the characteristic function for the exponentially modified Gaussian PDF (EMG).  The PDF that corresponds to the EMG is given below. 
\begin{eqnarray}
    P_{EMG}(x) = && {c_0\over 2} e^{\frac{c^2_0\sigma^2_0}{2}} 
     e^{-c_0(x-q_0)} \nonumber \\  
     && \times \textit{erfc}\left[ {1\over \sqrt{2}} \left( c_0\sigma_0 - {{x-q_0}\over \sigma_0}\right)  \right] 
    \label{eq:egpdf} 
\end{eqnarray}
$\textit{erfc}(x)$ is the complementary error function defined as $\textit{erfc}(x)= 1- \textit{erf}(x)$.  Some care is needed when evaluating this form of the PDF when the exponential parameter is too small or too large, but there are alternative forms that allow the evaluation.   We will not dwell on these details here.  In the following we will use definitions for the normal and the EMG PDFs as follows. $N(x; \mu, \sigma)$ is defined to be the normal PDF with mean of $\mu$ and standard deviation of $\sigma$.  $E_N(x; \mu,\sigma, c_0)$ is defined as the exponentially modified Gaussian PDF with $c_0$ as the additional exponential parameter. In the following we show that with these two PDFs, we can provide a detailed model for the charge from a photo-multiplier.  
In figure \ref{fg:emg}  we show examples of the function 
$E_N(x; \mu, \sigma, c_0)$. It is easily seen that when 
$\sigma$ is small the $\textit{erfc}$ function approaches 
a step function and the overall PDF approaches
an exponential.  A simple exercise will show that the 
mean of the exponentially modified Gaussian is $\mu+1/c_0$
and the variance is $\sigma^2+1/c_0^2$, with parameters as defined in this paragraph.  The mean is obviously shifted by the presence of the exponential, and the variance becomes larger because of the tail introduced by the exponential.  

\subsubsection{\label{sec:sig}Signal Charge}  

We now turn our attention to the probability density function of the signal charge. If $\lambda$ is the mean number of photo-electrons produced at the photo-cathode then the discrete random variable $K$ photo-electrons has the probability mass function given by the Poisson formula; $P_K(k)=e^{-\lambda}\lambda^k/k!$.  This Poisson probability distribution is  well known to come from a convolution of  Poisson and binomial processes 
when PMTs are used to sample low intensity sources.  
Each of the produced photo-electrons will go through the multiplication stage. The fluctuations of the gain or multiplication, $L$,  are dominated by the gain at the first stage due to secondary emission of electrons.   The mean first stage gain could range from zero to a few electrons depending on the 
 applied voltage; this gain is also Poisson distributed because only a small fraction of the electrons excited 
 in the dynode escape and are collected to contribute to the gain \cite{akkerman}.  
 If the mean gain at the first stage is $\alpha$ then the first stage gain is given by 
 a Poisson random variable : $P_L(l)=e^{-\alpha}\alpha^l/l!$.  
 The total charge from the signal process is the random variable $Z= \sum_{i=1}^K L_i$, where $L_i$ are independent random 
 numbers corresponding to the gain for each of the $K$ photo-electrons.   
The total charge, $Z$,  is recognized to be distributed according to the compound-Poisson\cite{feller, springer} probability function with a jump distribution that is also Poisson. A {\it jump distribution} is defined as the distribution that governs the statistics of the gain or $L$ in a compound Poisson probability distribution. 
In the case of a Poisson jump distribution, the charge $Z$ is a non-negative integer and the compound-Poisson is a discrete probability mass function.  In most ordinary circumstances, $\alpha$ is reasonably large, and  the gain can be characterized as normally distributed with a mean $\mu$ and standard deviation $\sigma$.  We have found that for values of $\alpha > 2$, the error made by assuming a Gaussian gain (with $\mu = \alpha$ and $\sigma = \sqrt{\alpha}$) is  $< 5\%$, adequate for most work\cite{nssc}.  Usually 
photo-multipliers run with first stage gain of $>4$, and therefore the assumption of Gaussian gain is quite sensible.  For the purposes of this paper, we will provide new formulas for both Poisson and Gaussian gain models. It is again easiest to start with the characteristic 
functions for the compound-Poisson probability functions. 
For a Poisson jump distribution, the characteristic function for $Z$, the signal charge is given as: 
\begin{eqnarray}  
\phi_Z(s) = && \textit{Exp}\left(  \lambda (e^{\alpha (e^{is}-1)} -1)     \right) \nonumber \\ 
= &&  \sum_{k=0}^\infty e^{k \alpha (e^{is}-1)} \times \frac{e^{-\lambda} \lambda^k}{k!}
\label{eq:cpp} 
\end{eqnarray}
And the discrete probability mass function is given by 
\begin{eqnarray}  
P_Z(n) =\begin{cases} 
e^{-\lambda} +  \sum_{k=1}^\infty e^{-k\alpha} \frac{e^{-\lambda} \lambda^k}{k!}  ~~~~ \dots  (n=0)  \nonumber \\ 
\\
  \sum_{k=1}^\infty \frac{e^{-k\alpha} (k\alpha)^n}{n!}  \frac{e^{-\lambda} \lambda^k}{k!} ~~~~ \dots  (n>0) 
  \end{cases}  
  \\
\label{eq:cpp2} 
\end{eqnarray}
For a Gaussian jump distribution,  we also provide the characteristic function and the probability density function for $Z$, however it should be remarked that in this case the probability density function is for a charge that is a continuous variable, and it can take negative values.  The characteristic function  for compound Poisson with Gaussian jump is given by:
\begin{eqnarray}
\phi_Z(s) =  \textit{Exp}\left(
\lambda (e^{is\mu - s^2\sigma^2/2}-1) 
\right) 
= \sum_{k=0}^\infty e^{isk\mu}  e^{-s^2\sigma^2 k\over 2} 
{e^{-\lambda} \lambda^k\over k!} 
\label{eq:cpg} 
\end{eqnarray}
The corresponding probability density function is obtained by noticing that each of the s-dependent terms above corresponds to a Gaussian PDF, except at $k=0$.  The $k=0$ term obviously can only contribute to no charge, and so we have to separate the probability density at zero. When we consider the total 
charge including the background this issue will disappear, but for completeness we provide the PDF for the signal only:  
\begin{eqnarray} 
P_Z(x) = \begin{cases} 
e^{-\lambda} + \sum_{k=1}^\infty {e^{-\lambda} \lambda^k\over k!} \times {e^{-k\mu^2/2\sigma^2}\over \sqrt{2\pi k \sigma^2} } ~~~~ \dots  \textit{for } (x=0) \nonumber \\  \\
\sum_{k=1}^\infty {e^{-\lambda} \lambda^k\over k!} \times {e^{-\frac{(x-k\mu)^2} {2k\sigma^2}}\over \sqrt{2\pi k \sigma^2} } ~~~~~ \dots  \textit{for }(x > 0) 
\end{cases}  
\\
\label{eq:cpg2} 
\end{eqnarray} 
Before ending this section, it is useful to consider the mean and the variance of the above signal charge distributions.  
These can be readily calculated by differentiating the characteristic function, $\phi_Z$ at $s=0$,  and multiplying with appropriate factors of $i$.  Obviously,  in the case  of  Poisson gain distribution with a parameter of $\alpha$, the mean signal will be $\lambda \alpha$,  and its variance will be $\lambda \alpha (\alpha+1) $. For normally distributed gain with a mean gain of $\mu$ and standard deviation of $\sigma$, the mean for the signal will be $\lambda \mu$ and the  variance  $\lambda(\sigma^2+ \mu^2)$. These results can be found in textbooks on probability and statistics \cite{feller}.  

\subsubsection{\label{sec:totchar}Total Charge}  
Now we must combine the signal and background charges to obtain the probability density function for the total charge.  We define the random variable, $Y=Z+B$, for the charge collected at the anode of the PMT.  The probability density function for $Y$ will be a convolution of the densities for $Z$ and $B$.  
 For the calculation in this section we will assume that the gain per photo-electron is best described by a Gaussian PDF.  In most ordinary circumstances this will be the case; the corresponding formula for a Poisson distributed gain will be  covered in the appendix.  The characteristic function for the total charge $Y$ with the assumption of the normal PDF for the gain is given by simply multiplying the functions from equations \ref{eq:bchar} and \ref{eq:cpg}.  
\begin{eqnarray} 
\phi_Y(s) = && \phi_Z(s)\times  \phi_D(s) \times \phi_Q(s)  \nonumber \\  
=  &&  
\left[ \sum_{k=0}^\infty e^{isk\mu}  e^{-s^2\sigma^2 k\over 2} 
{e^{-\lambda} \lambda^k\over k!}    \right]\times 
\left( (1-w) + w \frac{1}{1-is/c_0}  \right)\times  e^{isq_0}e^{-\frac{1}{2} \sigma^2_0 s^2} \nonumber \\ 
= &&  
\left[ \sum_{k=0}^\infty e^{is(k\mu+q_0)}  e^{-s^2(\sigma^2 k+\sigma_0^2)\over 2} 
{e^{-\lambda} \lambda^k\over k!}    \right]\times 
\left( (1-w) + w \frac{1}{1-is/c_0}  \right)
\label{eq:ychar}  
\end{eqnarray} 
Upon inspection we see that the terms multiplying $(1-w)$ will have a normal PDF and the terms multiplying $w/(1-is/c_0)$ will have the exponentially modified Gaussian PDF.  For convenience we provide the full PDF separated for no photo-electron, $k=0$, single photo-electron, $k=1$, and multiple photo-electrons.
We use the previous definitions for the normal and the EMG PDFs from section \ref{sec:back}.  Also recall that the form below is now valid for the entire 
domain of the real random variable $Y$ from negative to positive values.  
\begin{eqnarray} 
P_Y(x) = &&  e^{-\lambda}  \times \left(  (1-w)N(x;q_0,\sigma_0) +  w E_N(x;q_0, \sigma_0, c_0)  \right)   \nonumber \\ 
+ && \lambda e^{-\lambda}  \times \left( 
(1-w)N(x;\mu+q_0,\sqrt{\sigma^2+\sigma^2_0}) +  w E_N(x;\mu+q_0, \sqrt{\sigma^2+\sigma^2_0}, c_0)
\right)   \nonumber \\  
+ && \sum_{k=2}^\infty  \frac{\lambda^k e^{-\lambda}}{k!}   \times  \nonumber \\ 
&& \left( 
(1-w)N(x;k\mu +q_0,\sqrt{k\sigma^2+\sigma^2_0}) +  w E_N(x;k\mu+q_0, \sqrt{k\sigma^2+\sigma^2_0}, c_0)
\right) 
\label{eq:ypdf} 
\end{eqnarray} 

\section{\label{sec:lim} Limiting Forms } 
\begin{figure}[t]
\includegraphics[width=0.9\textwidth]{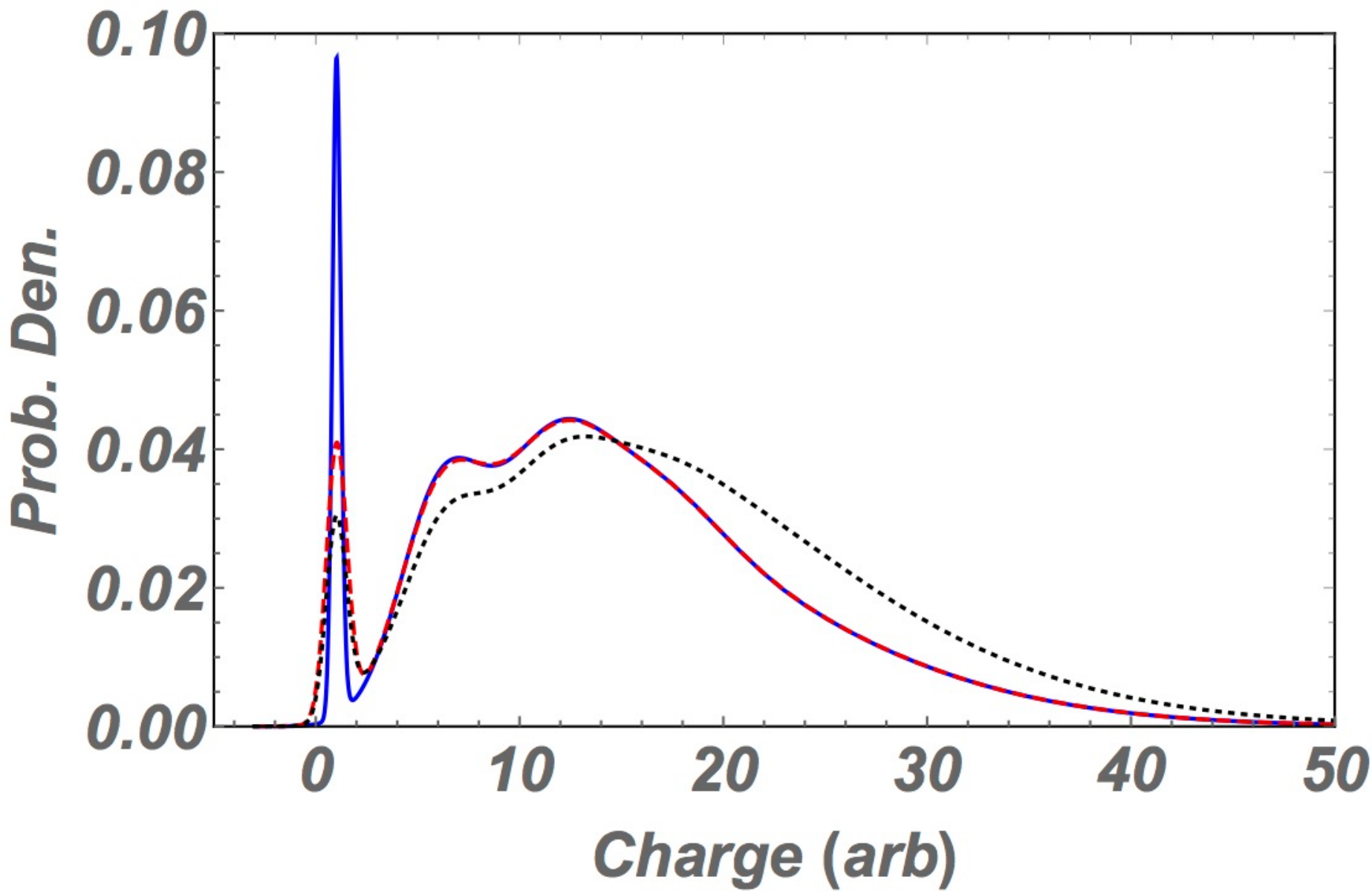}
\caption{\label{fg:spec1} 
Examples for the calculation of a PMT spectrum. The probability  of dark pulses, $w=0.3$.  
The values of parameters are 
set to $\lambda=3$ (mean number of photo-electrons),  $q_0=1$ (baseline shift), $\mu=5$ (mean gain), $\sigma = 2$ (standard deviation of the gain), $c_0=10$ (dark rate parameter),  and $\sigma_0 = 0.2 $   for the blue curve. For the red dashed curve we change the 
baseline fluctuation to 
$\sigma_0 = 0.5$.  And for the black dotted curve we change the dark rate parameter $c_0=1/\mu=1/5$ to  show its detrimental effect.  
Background processes do not simply add to the spectrum, but they change the spectra due to the convolutions with signal processes.  
}
\end{figure}

\begin{figure}[t]
\includegraphics[width=0.9\textwidth]{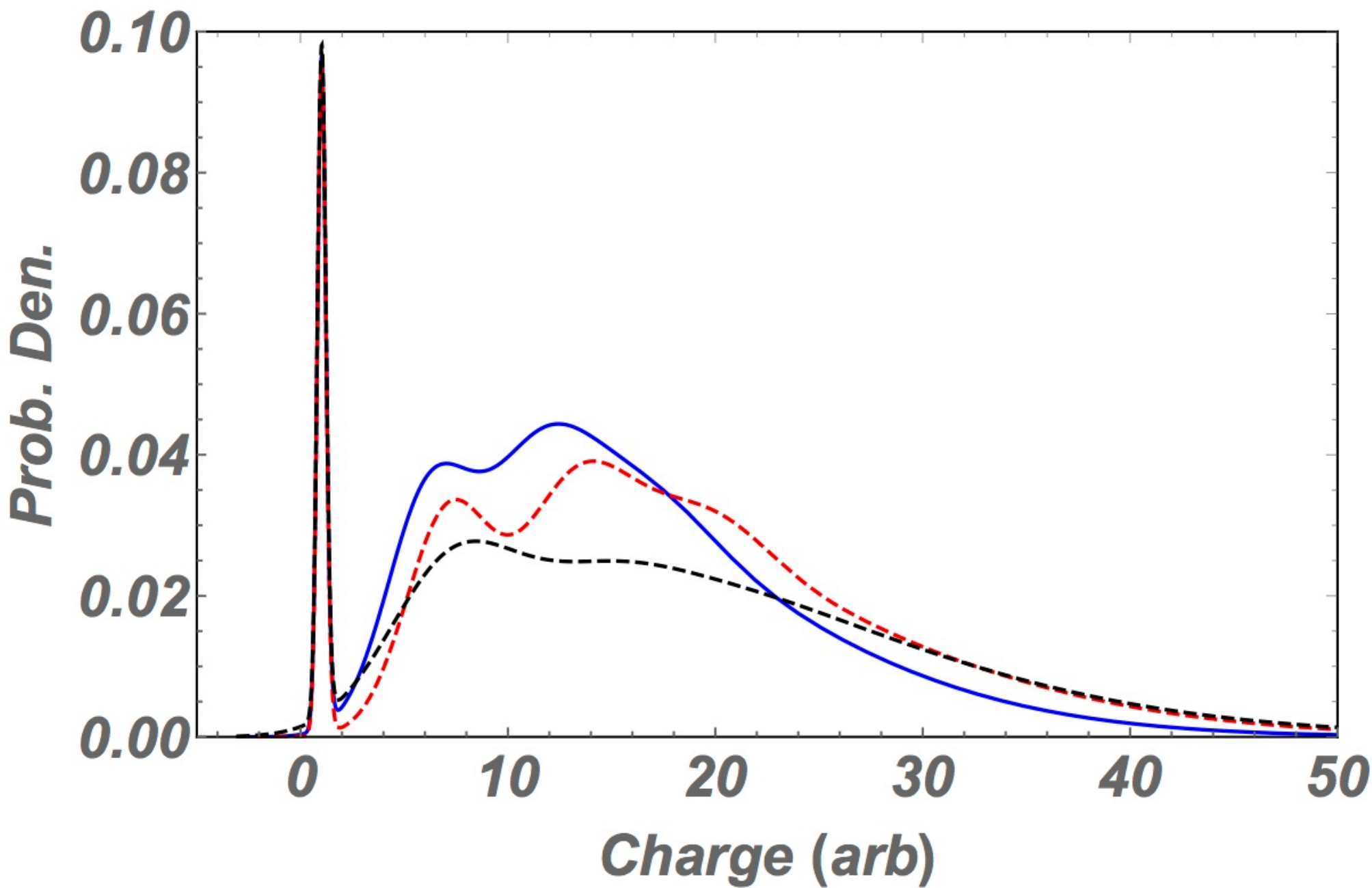}
\caption{\label{fg:spec2} 
Examples for the calculation of  PMT spectra and its dependence on the gain parameters.  
The probability of dark pulses, $w=0.3$.  
The values of parameters are 
set to be $\lambda=3$ (mean photo-electrons),  $q_0=1$ (baseline shift), $\sigma_0=0.2$ (baseline fluctuation), $c_0=10$ (dark rate parameter), $\mu=5$ (mean gain), $\sigma =2$ (gain fluctuation)   for the blue curve.  
For the red dashed curve we change the mean gain $\mu = 6$ which clearly shows as a shift in the spectra.  
For the black dashed curve we additionally change the gain fluctuation parameter $\sigma = 3$. 
}
\end{figure}

We first examine the spectra resulting from equation \ref{eq:ypdf} in figures \ref{fg:spec1}, and \ref{fg:spec2}.  
Figure \ref{fg:spec1} shows the effect of the baseline 
fluctuations. As the baseline fluctuation grows, the pedestal will become wider, however the effect on the main part of the spectrum remains small as long as $\sigma_0 \ll \sigma$. The effect of dark pulses on the spectrum is small, even with $w=0.3$  as long as $c_0$ remains large. 

If the baseline fluctuation is small so that $c_0 \sigma_0 \ll 1$
then the first line in equation \ref{eq:ypdf} becomes a simple sum of a pedestal and an exponential background \cite{bellamy}:
\begin{eqnarray} 
P_{B}(x) \approx  (1-w)N(x;q_0,\sigma_0)+w\theta(x)c_0 e^{-c_0(x-q_0)}.\nonumber \\
\label{eq:bonly} 
\end{eqnarray}  
For a large $c_0$ the dark rate has small pulse heights, and therefore the effect on the main part of the spectrum is small but also for very small $c_0$ the dark rate does not affect the shape of the main part of the spectrum because the dark pulses become spread out over a large range of charge,  diminishing their contribution. Of course, one expects the dark pulses to contribute pulse heights that are in the same range as the single photo-electron charge, $c_0\sim 1/\mu$, where they have the maximum detrimental effect on the measurement. This can be seen in figure \ref{fg:spec1}.   If the dark rate characteristics are such that it can be neglected from the signal part of the spectrum, then the spectrum can be simplified to have separate contributions from pedestal, dark rate, and signal.  However this can be an over-simplification in many circumstances. Here we also make the assumption that the  baseline fluctuations are small compared to the gain fluctuations, $\sigma_0\ll \sigma $, but the baseline shift ($q_0$) is retained in the equation:  
\begin{eqnarray} 
P_Y(x) \approx &&  e^{-\lambda}  \times \left(  (1-w)N(x;q_0,\sigma_0) +  w \theta(x) c_0 e^{-c_0 (x-q_0) }   \right)   \nonumber \\ 
+ && \sum_{k=1}^\infty  \frac{\lambda^k e^{-\lambda}}{k!}   \times
N(x;k\mu +q_0,\sqrt{k\sigma^2}).
\end{eqnarray}  

In figure \ref{fg:spec2} we show a comparison between
spectra with differing values of the gain ($\mu$ ) and the gain fluctuation ($\sigma$).  It is clear that for low light levels,
the mean number of photo-electrons can be obtained from the distinctive shape of the spectrum. Ideally, if the baseline 
fluctuation is small,  the total probability content around the pedestal can be measured to obtain $(1-w) e^{-\lambda} $. For small dark current, this will provide a simple estimate of the 
mean number of photo-electrons.  A subsequent fit to the spectrum can be made to fit $\lambda, \mu$, and  $\sigma$.  

\begin{figure}[t]
\includegraphics[scale=0.8]{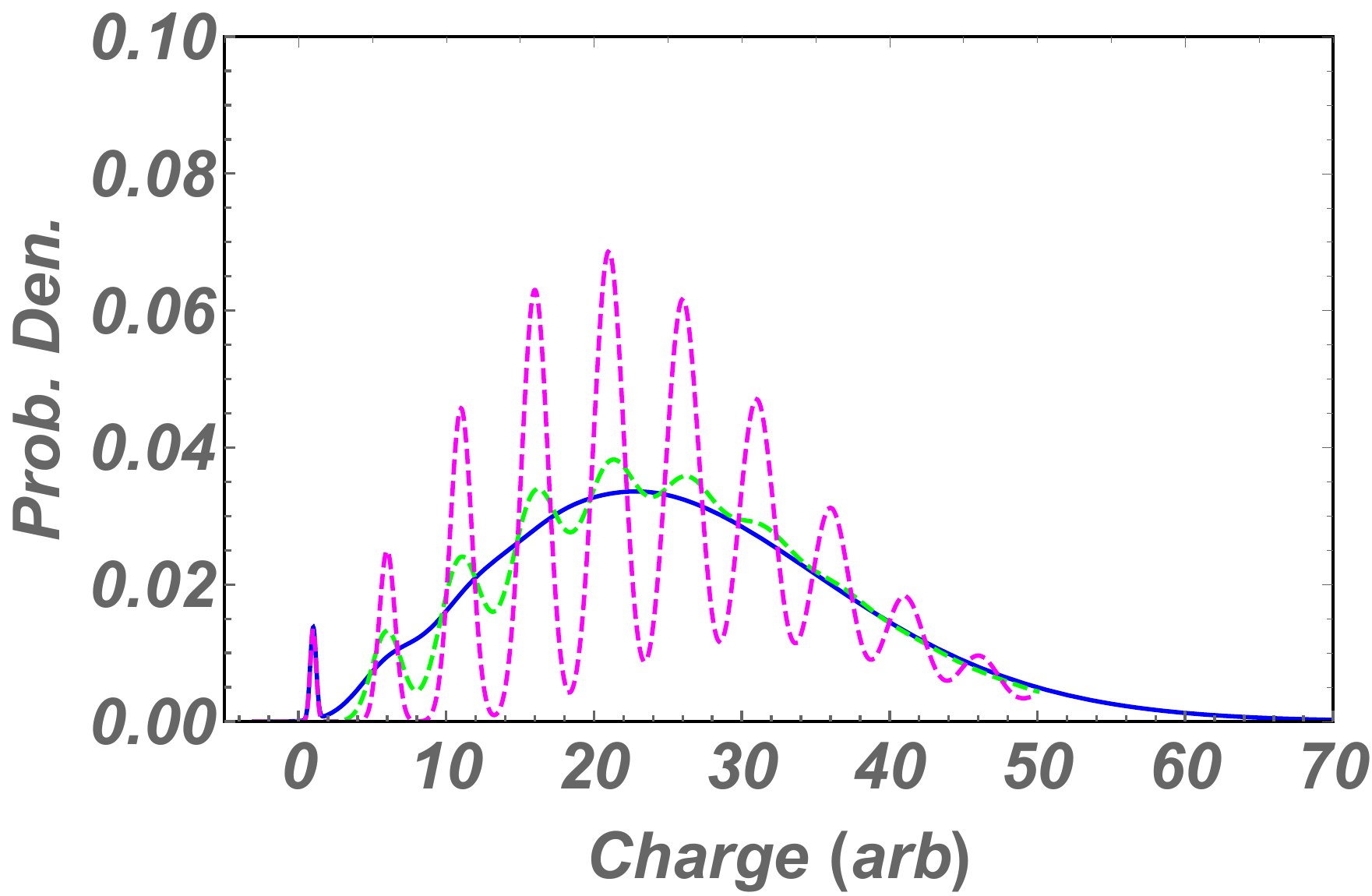}
\caption{\label{fg:spec3} 
Examples of PMT spectra with differing gain fluctuations.  
The probability of dark pulses is set to zero, $w=0$.
The values of other parameters are 
set to be $\lambda=5$ (mean photo-electrons),  $q_0=1$ (baseline shift), $\sigma_0=0.2$ (baseline fluctuation),  $\mu=5$ (mean gain), $\sigma =0.5, 1, 2$ (gain fluctuation)   for the magenta-dashed, green-dashed, and blue-solid  curves, respectively.  
}
\end{figure}

When the gain fluctuations are small compared to the fluctuations in the number of photo-electrons, $\sigma/\mu \ll 1/\sqrt{\lambda}$,  the signal probability density becomes a sum of narrow Gaussian peaks separated by the mean value of the gain, $\mu$.
In such a case, the mean photo-electron count, as well as the gain can be measured from the spectrum with little difficulty.  However, if the gain fluctuations are comparable to the fluctuations due to the number of photo-electrons, the two are difficult to separate from each other. 
When $\lambda$, the average number of photo-electrons becomes large and the gain fluctuations are comparable to $1/\sqrt{\lambda}$,  the signal probability density (compound Poisson) can be approximated as  Gaussian (figure \ref{fg:spec3}).
This is sometimes erroneously described as a convolution of Poisson and Gaussian distributions. 
The form of the approximate signal Gaussian when the gain fluctuations are comparable to $1/\sqrt{\lambda}$ is as follows:  
\begin{eqnarray} 
P_Z(x) \approx N(x; \lambda\mu, \sqrt{\lambda(\sigma^2+\mu^2))} 
\label{eq:lrglambda}  
\end{eqnarray} 
In this case, it is likely safe to neglect the background due to baseline 
fluctuations; the spectrum after combining with the background becomes: 
\begin{eqnarray} 
P_Y(x) \approx (1-w) N(x; \lambda\mu+q_0, \sqrt{\lambda (\sigma^2+\mu^2))} + 
w E_N(x; \lambda\mu+q_0, \sqrt{\lambda (\sigma^2+\mu^2)}, c_0) 
\label{eq:lrglspec}  
\end{eqnarray} 
Here we retain the baseline shift, but ignore the contribution from the baseline fluctuation.  
The  overall  mean of the spectrum, 
$(\lambda \mu + q_0 + w/c_0 )$, 
will appear  further shifted if  the dark pulse contribution is sizable.  

\section{\label{sec:dis} Discussion }

\begin{figure}[t]
\includegraphics[scale=0.7]{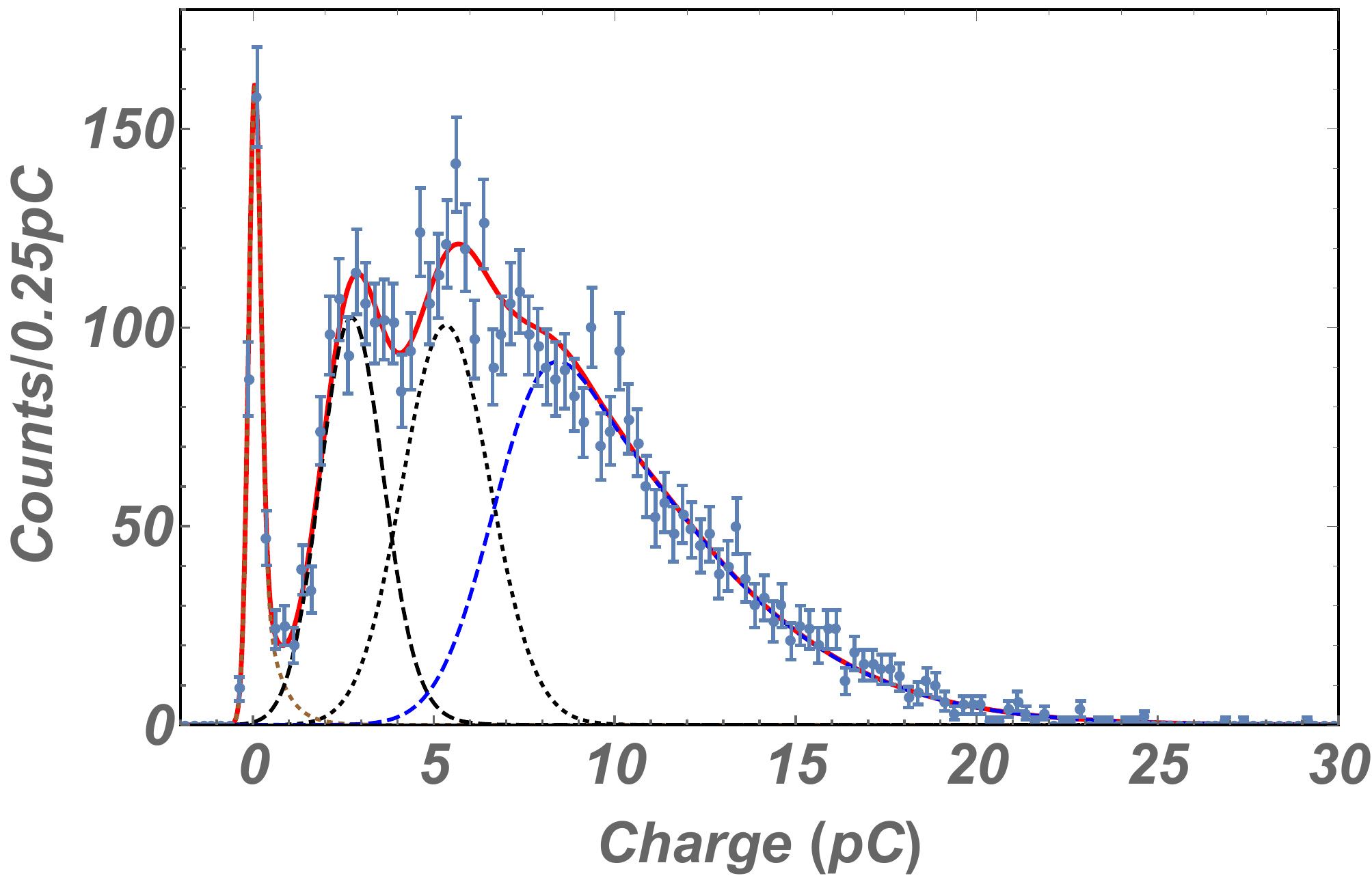}
\caption{\label{fg:data} 
Example fit to data (blue dots)  using equation \ref{eq:ypdf}. The data contains  5000 pulses. The 7 parameter 
fit resulted in a reduced $\chi^2/DOF=1.08$ for 77 DOF.  
The parameters extracted were the baseline shift $q_0=0.044\pm0.031$pC,  the baseline fluctuation $\sigma_0= 0.169\pm 0.018$pC,
the dark rate probability $w=0.30\pm 0.12$,  the dark rate exponential parameter $c_0=2.1\pm2.1 {\rm pC^{-1}} $,  the mean gain $\mu = 2.59\pm 0.06$pC, 
the gain fluctuation $\sigma = 0.826\pm 0.057$pC and mean number of photo-electrons $\lambda = 2.69\pm0.10$ . Please see the text for further comments.   The figure shows the best fit curve (red) as well as individual components of the spectrum: the charge spectrum for no photo-electron emission (brown dashed) shows a small dark rate component as a tail on the positive side; the single photo-electron spectrum (black dashed), a two photo-electron spectrum (black dotted), and greater than two photo-electrons (blue dashed) are shown separately.  
}
\end{figure}

The formula derived  in equation \ref{eq:ypdf}  can be used for fitting and extracting calibration constants for photo-multipliers in various circumstances.  
A thorough analysis of fitting with data is beyond the scope of this paper which is focused on the derivation and analysis of the underlying 
distributions, nevertheless we have provided an example of data and a fit using \ref{eq:ypdf} in figure \ref{fg:data}.  The data was obtained by 
flashing a low intensity light pulse on a Hamamatsu R5912 photomultiplier and read out using a Series-6 Tektronics oscilloscope 
using an external trigger. 
We now provide a partial list of considerations when using the formula and the methods described above: 
\begin{itemize} 
\item  For purposes of gain calibration and 
equalization of gains,  it is common to have a light pulser system for PMTs. 
The best way to use this is to pulse with low intensity (few photo-electrons) with pulse widths that are smaller than the 
fastest time constant for the PMT (see figure \ref{fg:data}).  It is not necessary to have the same intensity for all channels. The formula provided above shows that for low intensity, the gain and the number of photo-electrons can be disentangled very well.    
The example shown in figure \ref{fg:data} demonstrates that equation \ref{eq:ypdf}  describes the entire spectrum including the pedestal in detail. 
The inclusion of the pedestal in the fit  is very useful to assure normalization of the spectrum,  and 
extract the mean number of photo-electrons, the baseline, and the electronic noise that 
contributes to the baseline fluctuations.  

\item It is important to distinguish between data acquired 
by utilizing a trigger on the channel that is being calibrated and by using a time window that is imposed externally. In the latter case, the collected charge will be unbiased and the probability content of the spectrum will be as described above. If a trigger is imposed on the data, then obviously the collected charge distribution will be biased. If the trigger threshold is accurate and there is little noise, then there will be a lower cutoff of the spectrum, but this is rarely the case.    A trigger can be considered  a non-linear filter that is imposed on the data, either in hardware or in software.  

\item  It is common for modern data acquisition systems to 
perform  waveform digitization with fast ADCs.  The calibration using the above formula would still work the same
as long as the data was acquired without bias.  An important consideration in the acquisition should be the bandwidth and the corresponding noise level.

\item The formula is not limited to very constrained data acquisition conditions such as a fixed time window of acquisition and fixed light intensity from a pulser.  As long as the data acquired is unbiased, and the conditions are not changing in time, the formula can be used to extract the mean number of photo-electrons and the mean gain.  Such conditions could result from a constant rate radioactive decay or cosmic ray muons in a scintillating medium chosen to provide  calibration data during a long data-taking run.    The data could be acquired using a long random time window or triggered using external detectors.
We recommend statistics of at least ~5000 pulse acquisitions to provide  good accuracy.  

\item If calibration pulser data cannot be low intensity ($\lambda < 10 $ photo-electrons) and the gain fluctuations are large, then we are in a regime where the average number of photo-electrons and the gain cannot be easily separated from each other.    There are strategies that could be employed by obtaining spectra at several high voltage settings.  These strategies rely on the observation that the gain and the variance of the gain 
are related to each other through the amplification at the first
dynode. If the voltage drop at the first dynode is fixed (by the use of a Zener diode, as an example) then the gain fluctuation remains 
constant, and the ratio of the standard deviation of the gain and the gain should remain the same ($\sigma/\mu$) .  
But if the voltage at the first dynode is linear with the overall voltage at the cathode, then the ratio of the gain and standard deviation of the gain will vary as the square-root of the first stage gain. These relationships could be employed in the fit to constrain the gain parameters with respect to voltage to separate the 
average intensity of the photo-electrons.  

\item With the advent of waveform digitization, the emphasis
of simulation is often on simulating the waveform. We find this unnecessary for most low rate applications (such as neutrino physics).  The expected number of photo-electrons can be simulated using geometric and optical simulations for each PMT. The distribution function using calibrated constants and the integral of the formula in equation \ref{eq:ypdf} can then be used to directly simulate the measured charge with the correct model for 
experimental fluctuations.  

\item We make a final remark regarding the number of terms that should be retained in equation \ref{eq:ypdf} for accuracy in 
numerical calculations.  This obviously depends on $\lambda$, the mean number of photo-electrons. We suggest the sum should be carried out to at least 5 to 10 times $\lambda$ depending on the
needed accuracy and computing speed.  

\end{itemize}

\section{\label{sec:con} Conclusion}  

We have provided a new detailed formula for the charge spectrum of a photo-multiplier tube subjected to low light illumination. The only assumptions in this exercise were that the  response of the PMT is independent for each produced photo-electron, and that the charge is collected in an unbiased way.  We suggest the use of this formula for fitting and calibrating experimental spectra, and also for accurate fast Monte Carlo generation in complex experimental situations.  In the derivation of this result a number of improvements have been made with respect to existing literature. 
A general formalism is provided with broad applicability; it can be used for a variety of detectors (e.g. silicon photomultipliers) 
as well as statistical models for signal, 
background, and gain.  An application for photo-multiplier tubes is thoroughly investigated including an example fit to data.  
A new formula using Poisson gain model is  provided in the appendix.  The new analysis does not require the assumption of low charge 
for background processes which affect the spectra. This  analysis also allows precise determination of the probability density 
at and near zero charge which can occur due to the  sum of fluctuations in charges from backgrounds and signal.

\begin{acknowledgments}
This work was supported by the US Department of Energy.  It was presented at Nuclear Science and Security Consortium Summer School 2019 at the University of California/Davis. I thank Bob Svoboda for encouraging me to write it up. 
\end{acknowledgments}

\appendix

\section{Alternate Gain Model}

For completeness we provide the full expression (equivalent to 
equation \ref{eq:ypdf}), but for an assumption of gain distributed as if a Poisson distribution. When the first stage gain is sufficiently large the equation can easily be shown to approach the one for Gaussian gain.  In this formulation, many elementary Gaussian and exponential-modified-Gaussians shifted to the corresponding gain are seen to combine with appropriate weights to sum up to the total spectrum.  

\begin{eqnarray} 
P_Y(x) = &&  e^{-\lambda}  \times \left(  (1-w)N(x;q_0,\sigma_0) +  w E_N(x;q_0, \sigma_0, c_0)  \right)   \nonumber \\ 
+ && \sum_{n=0}^\infty  \lambda e^{-\lambda} \frac{(\alpha)^n e^{-\alpha}}{n!} \times \left( 
(1-w)N(x;n+q_0,\sigma_0) +  w E_N(x;n+q_0, \sigma_0, c_0)
\right)   \nonumber \\  
+ && \sum_{k=2}^\infty \sum_{n=0}^\infty \frac{\lambda^k e^{-\lambda}}{k!} \frac{(k\alpha)^n e^{-k\alpha}}{n!} \times 
\left( 
(1-w)N(x;n+q_0,\sigma_0) +  w E_N(x;n+q_0, \sigma_0, c_0)
\right) 
\end{eqnarray}  

\bibliography{photomult-spect-jinst}{}

\providecommand{\noopsort}[1]{}\providecommand{\singleletter}[1]{#1}%

\providecommand{\href}[2]{#2}\begingroup\raggedright\begin{thebibliography}{10}

\bibitem{knoll}
G.~F. Knoll, \emph{Radiation Detection and Measurement}.
\newblock Wiley, 2010.

\bibitem{photonis}
S.-O. Flyct and C.~Marmonier, \emph{Photomultiplier Tubes, Principles and
  Applications}.
\newblock Photonis, 2002.

\bibitem{boger:1999bb}
J.~Boger et~al., \emph{The sudbury neutrino observatory}, {\emph{Nucl. Inst.
  Meth.} {\bfseries A} (2000) }.

\bibitem{fukuda:2002uc}
Y.~Fukuda et~al., \emph{The super-kamiokande detector}, {\emph{Nucl. Inst.
  Meth.} {\bfseries A} (2003) }.

\bibitem{bellamy}
E.~Bellamy et~al.{\emph{Nucl. Inst. Meth.} {\bfseries A339} (1994) 468}.

\bibitem{pdg}
M.~Tanabashi et~al., \emph{Particle data group}, {\emph{Phys. Rev. D}
  {\bfseries 98} (2018) }.

\bibitem{akkerman}
A.~Akkerman et~al., \emph{Low-energy electron transport in alkali halides},
  {\emph{Journal of Applied Physics} {\bfseries 76} (1994) }.

\bibitem{feller}
W.~Feller, \emph{Introduction to Probability Theory and Its Applications}.
\newblock John Wiley and Sons, 1960.

\bibitem{springer}
M.~D. Springer, \emph{The Algebra of Random Variables}.
\newblock Wiley, 1979.

\bibitem{nssc}
M.~Diwan, \emph{Photomultiplier spectra and absolute calibration},  Nuclear
  Analytics Techniques Summer School, (University of California/Davis), Nuclear
  Analytics Techniques Summer School, 2019.

\end{thebibliography}\endgroup
\bibliographystyle{JHEP}  

\end{document}